\documentclass[modern]{aastex62}

\newcommand{\docRef}{set the Reference with {$\backslash$}setDocRef}
\newcommand{\setDocRef}[1]{
   \renewcommand{\docRef}{#1}
}

\setDocRef{DMTN-115}
\usepackage{newtxtext,newtxmath} 
\usepackage[USenglish]{babel}
\usepackage[utf8]{inputenc}
\usepackage[T1]{fontenc}
\usepackage{comment}
\usepackage{enumitem}
\usepackage{glossaries}
\usepackage{todonotes}
\usepackage{collect}
\usepackage{wrapfig}

\setlist{noitemsep}

\graphicspath{{./}{./figures/}{./images}}

\providecommand{\figref}[1]{Figure~\ref{#1}}

\newacronym{API} {API} {Application Programming Interface}
\newglossaryentry{AURA} {name={AURA}, description={\gls{Association of Universities for Research in Astronomy}}}
\newacronym{AWS} {AWS} {Amazon Web Services}
\newglossaryentry{Alert} {name={Alert}, description={A packet of information for each source detected with signal-to-noise ratio > 5 in a difference image during Prompt Processing, containing measurement and characterization parameters based on the past 12 months of LSST observations plus small cutouts of the single-visit, template, and difference images, distributed via the internet.}}
\newglossaryentry{Alert Production} {name={Alert Production}, description={The principal component of Prompt Processing that processes and calibrates incoming images, performs Difference Image Analysis to identify DIASources and DIAObjects, packages and distributes the resulting Alerts, and runs the Moving Object Processing System.}}
\newglossaryentry{Archive} {name={Archive}, description={The repository for documents required by the NSF to be kept. These include documents related to design and development, construction, integration, test, and operations of the LSST observatory system. The archive is maintained using the enterprise content management system DocuShare, which is accessible through a link on the project website www.project.lsst.org.}}
\newglossaryentry{Archive Center} {name={Archive Center}, description={Part of the LSST Data Management System, the LSST archive center is a data center at NCSA that hosts the LSST Archive, which includes released science data and metadata, observatory and engineering data, and supporting software such as the LSST Software Stack.}}
\newglossaryentry{Association Pipeline} {name={Association Pipeline}, description={An application that matches detected Sources or DIASources or generated Objects to an existing catalog of Objects, producing a (possibly many-to-many) set of associations and a list of unassociated inputs. Association Pipelines are used in Prompt Processing after DIASource generation and in the final stages of Data Release processing to ensure continuity of Object identifiers.}}
\newglossaryentry{Association of Universities for Research in Astronomy} {name={Association of Universities for Research in Astronomy}, description={ consortium of US institutions and international affiliates that operates world-class astronomical observatories, AURA is the legal entity responsible for managing what it calls independent operating Centers, including LSST, under respective cooperative agreements with the National Science Foundation. AURA assumes fiducial responsibility for the funds provided through those cooperative agreements. AURA also is the legal owner of the AURA Observatory properties in Chile.}}
\newacronym{CAOM} {CAOM} {Common Archive Observation Model}
\newacronym{CI} {CI} {Cyber Infrastructure}
\newglossaryentry{CRUD} {name={CRUD}, description={Create Retrieve Update and Destroy}}
\newacronym{CSV} {CSV} {Comma Separated Values}
\newglossaryentry{Center} {name={Center}, description={An entity managed by AURA that is responsible for execution of a federally funded project}}
\newacronym{DCR} {DCR} {Differential Chromatic Refraction}
\newacronym{DIA} {DIA} {Difference Image Analysis}
\newglossaryentry{DIAObject} {name={DIAObject}, description={A DIAObject is the association of DIASources, by coordinate, that have been detected with signal-to-noise ratio greater than 5 in at least one difference image. It is distinguished from a regular Object in that its brightness varies in time, and from a SSObject in that it is stationary (non-moving).}}
\newglossaryentry{DIASource} {name={DIASource}, description={A DIASource is a detection with signal-to-noise ratio greater than 5 in a difference image.}}
\newacronym{DM} {DM} {\gls{Data Management}}
\newacronym{DMS} {DMS} {Data Management Subsystem}
\newglossaryentry{DMTN} {name={DMTN}, description={DM Technical Note}}
\newacronym{DR} {DR} {Data Release}
\newacronym{DRP} {DRP} {Data Release Production}
\newglossaryentry{Data Management} {name={Data Management}, description={The LSST Subsystem responsible for the Data Management System (DMS), which will capture, store, catalog, and serve the LSST dataset to the scientific community and public. The DM team is responsible for the DMS architecture, applications, middleware, infrastructure, algorithms, and Observatory Network Design. DM is a distributed team working at LSST and partner institutions, with the DM Subsystem Manager located at LSST headquarters in Tucson.}}
\newglossaryentry{Data Management Subsystem} {name={Data Management Subsystem}, description={The subsystems within Data Management may contain a defined combination of hardware, a software stack, a set of running processes, and the people who manage them: they are a major component of the DM System operations. Examples include the 'Archive Operations Subsystem' and the 'Data Processing Subsystem'"."}}
\newglossaryentry{Data Management System} {name={Data Management System}, description={The computing infrastructure, middleware, and applications that process, store, and enable information extraction from the LSST dataset; the DMS will process peta-scale data volume, convert raw images into a faithful representation of the universe, and archive the results in a useful form. The infrastructure layer consists of the computing, storage, networking hardware, and system software. The middleware layer handles distributed processing, data access, user interface, and system operations services. The applications layer includes the data pipelines and the science data archives' products and services.}}
\newglossaryentry{Data Release} {name={Data Release}, description={The approximately annual reprocessing of all LSST data, and the installation of the resulting data products in the LSST Data Access Centers, which marks the start of the two-year proprietary period.}}
\newglossaryentry{Data Release Production} {name={Data Release Production}, description={An episode of (re)processing all of the accumulated LSST images, during which all output DR data products are generated. These episodes are planned to occur annually during the LSST survey, and the processing will be executed at the Archive Center. This includes Difference Imaging Analysis, generating deep Coadd Images, Source detection and association, creating Object and Solar System Object catalogs, and related metadata.}}
\newglossaryentry{Difference Image} {name={Difference Image}, description={Refers to the result formed from the pixel-by-pixel difference of two images of the sky, after warping to the same pixel grid, scaling to the same photometric response, matching to the same PSF shape, and applying a correction for Differential Chromatic Refraction. The pixels in a difference thus formed should be zero (apart from noise) except for sources that are new, or have changed in brightness or position. In the LSST context, the difference is generally taken between a visit image and template. }}
\newglossaryentry{Difference Image Analysis} {name={Difference Image Analysis}, description={The detection and characterization of sources in the Difference Image that are above a configurable threshold, done as part of Alert Generation Pipeline.}}
\newglossaryentry{Differential Chromatic Refraction} {name={Differential Chromatic Refraction}, description={The refraction of incident light by Earth's atmosphere causes the apparent position of objects to be shifted, and the size of this shift depends on both the wavelength of the source and its airmass at the time of observation. DCR corrections are done as a part of DIA.}}
\newglossaryentry{DocuShare} {name={DocuShare}, description={The trade name for the enterprise management software used by LSST to archive and manage documents}}
\newacronym{FITS} {FITS} {\gls{Flexible Image Transport System}}
\newglossaryentry{FSAAS} {name={FSAAS}, description={Filesystem as a Service}}
\newglossaryentry{FUSE} {name={FUSE}, description={a user space filesystem framework}}
\newglossaryentry{Flexible Image Transport System} {name={Flexible Image Transport System}, description={an international standard in astronomy for storing images, tables, and metadata in disk files. See the IAU FITS Standard for details.}}
\newacronym{HDF} {HDF} {Hierarchical Data Format}
\newacronym{HPC} {HPC} {High Performance Computing}
\newacronym{HTC} {HTC} {High Throughput Computing}
\newacronym{IAU} {IAU} {International Astronomical Union}
\newglossaryentry{IRAF} {name={IRAF}, description={Image Reduction and Analysis Facility}}
\newglossaryentry{JSON} {name={JSON}, description={JavaScript Object Notation}}
\newacronym{LSST} {LSST} {Large Synoptic Survey Telescope}
\newacronym{MOPS} {MOPS} {Moving Object Processing System}
\newglossaryentry{Moving Object Processing System} {name={Moving Object Processing System}, description={The Moving Object Processing System (MOPS) identifies new SSObjects using unassociated DIASources. MOPS is part of the Science Pipelines.}}
\newglossaryentry{NCSA} {name={NCSA}, description={National Center for Supercomputing Applications}}
\newacronym{NSF} {NSF} {\gls{National Science Foundation}}
\newglossaryentry{National Science Foundation} {name={National Science Foundation}, description={primary federal agency supporting research in all fields of fundamental science and engineering; NSF selects and funds projects through competitive, merit-based review}}
\newglossaryentry{Operations} {name={Operations}, description={The 10-year period following construction and commissioning during which the LSST Observatory conducts its survey}}
\newglossaryentry{POSIX} {name={POSIX}, description={Portable Operating System Interface}}
\newacronym{PSF} {PSF} {Point Spread Function}
\newglossaryentry{Project Manager} {name={Project Manager}, description={The person responsible for exercising leadership and oversight over the entire LSST project; he or she controls schedule, budget, and all contingency funds}}
\newglossaryentry{Prompt Processing} {name={Prompt Processing}, description={The processing that occurs at the Archive Center on the nightly stream of raw images coming from the telescope, including Difference Imaging Analysis, Alert Production, and the Moving Object Processing System. This processing generates Prompt Data Products.}}
\newglossaryentry{Science Pipelines} {name={Science Pipelines}, description={The library of software components and the algorithms and processing pipelines assembled from them that are being developed by DM to generate science-ready data products from LSST images. The Pipelines may be executed at scale as part of LSST Prompt or Data Release processing, or pieces of them may be used in a standalone mode or executed through the LSST Science Platform. The Science Pipelines are one component of the LSST Software Stack.}}
\newglossaryentry{Science Platform} {name={Science Platform}, description={A set of integrated web applications and services deployed at the LSST Data Access Centers (DACs) through which the scientific community will access, visualize, and perform next-to-the-data analysis of the LSST data products.}}
\newglossaryentry{Software Stack} {name={Software Stack}, description={Often referred to as the LSST Stack, or just The Stack, it is the collection of software written by the LSST Data Management Team to process, generate, and serve LSST images, transient alerts, and catalogs. The Stack includes the LSST Science Pipelines, as well as packages upon which the DM software depends. It is open source and publicly available.}}
\newglossaryentry{Solar System Object} {name={Solar System Object}, description={A solar system object is an astrophysical object that is identified as part of the Solar System: planets and their satellites, asteroids, comets, etc. This class of object had historically been referred to within the LSST Project as Moving Objects.}}
\newglossaryentry{Source} {name={Source}, description={A single detection of an astrophysical object in an image, the characteristics for which are stored in the Source Catalog of the DRP database. The association of Sources that are non-moving lead to Objects; the association of moving Sources leads to Solar System Objects. (Note that in non-LSST usage "source" is often used for what LSST calls an Object.)}}
\newglossaryentry{Subsystem} {name={Subsystem}, description={A set of elements comprising a system within the larger LSST system that is responsible for a key technical deliverable of the project.}}
\newglossaryentry{Subsystem Manager} {name={Subsystem Manager}, description={responsible manager for an LSST subsystem; he or she exercises authority, within prescribed limits and under scrutiny of the Project Manager, over the relevant subsystem's cost, schedule, and work plans}}
\newacronym{US} {US} {United States}
\newglossaryentry{airmass} {name={airmass}, description={The pathlength of light from an astrophysical source through the Earth's atmosphere. It is given approximately by sec z, where z is the angular distance from the zenith (the point directly overhead, where airmass = 1.0) to the source.}}
\newglossaryentry{algorithm} {name={algorithm}, description={A computational implementation of a calculation or some method of processing.}}
\newglossaryentry{astronomical object} {name={astronomical object}, description={A star, galaxy, asteroid, or other physical object of astronomical interest. Beware: in non-LSST usage, these are often known as sources.}}
\newglossaryentry{flux} {name={flux}, description={Shorthand for radiative flux, it is a measure of the transport of radiant energy per unit area per unit time. In astronomy this is usually expressed in cgs units: erg/cm2/s.}}
\newglossaryentry{metadata} {name={metadata}, description={General term for data about data, e.g., attributes of astronomical objects (e.g. images, sources, astroObjects, etc.) that are characteristics of the objects themselves, and facilitate the organization, preservation, and query of data sets. (E.g., a FITS header contains metadata).}}
\newglossaryentry{pipeline} {name={pipeline}, description={A configured sequence of software tasks (Stages) to process data and generate data products. Example: Association Pipeline.}}
\newglossaryentry{provenance} {name={provenance}, description={Information about how LSST images, Sources, and Objects were created (e.g., versions of pipelines, algorithmic components, or templates) and how to recreate them.}}
\newglossaryentry{shape} {name={shape}, description={In reference to a Source or Object, the shape is a functional characterization of its spatial intensity distribution, and the integral of the shape is the flux. Shape characterizations are a data product in the DIASource, DIAObject, Source, and Object catalogs.}}
\newglossaryentry{stack} {name={stack}, description={a grouping, usually in layers (hence stack), of software packages and services to achieve a common goal. Often providing a higher level set of end user oriented services and tools}}
\newglossaryentry{transient} {name={transient}, description={A transient source is one that has been detected on a difference image, but has not been associated with either an astronomical object or a solar system body.}}

\makeglossaries

\begin{document}

\author{William~O'Mullane.}
\affiliation{Large Synoptic Survey Telescope (LSST/AURA)}
\author{Niall~Gaffney}
\affiliation{Texas Advanced Computing Center}
\author{Frossie~Economou}
\affiliation{Large Synoptic Survey Telescope (LSST/AURA)}
\author{Arfon~M.~Smith}
\affiliation{Space Telescope Science Institute}
\author{J.~Ross~Thomson}
\affiliation{Google}
\author{Tim~Jenness}
\affiliation{Large Synoptic Survey Telescope (LSST/AURA)}

\date{\today}

\begin{abstract}
Many astronomy data centres still work on filesystems.
Industry has moved on;  current practice in  computing infrastructure is to achieve Big Data scalability using  object stores rather than POSIX file systems. This presents us with  opportunities  for portability and reuse of software underlying processing and archive  systems  but it also causes problems for legacy implementations in current data centers.

\end{abstract}

\title{The demise of the filesystem and multi level service  architecture}
 \hypersetup{pdftitle={\@title}, pdfauthor={\@author}, pdfkeywords={\docRef \@keywords}}

\section{Introduction} \label{sec:intro}

\textbf{Executive summary: the filesystem notion limits our ability to scale processing and has long since been dropped by industry. Astronomy needs to move on with a new architecture.}

Object Stores are nothing new to Astronomy.  \gls{FITS} and \gls{IRAF} tapes are examples
of object stores that were used when file systems were unable to handle the volumes
of data being produced. Even then, standards for migrating from Object Store to file
systems were created, allowing for objects to be retrieved into a predefined namespace
on disk.  The explosion of individual \gls{POSIX} disk capacity and \gls{POSIX}-like file systems
have produced generations of researchers who have never used an Object Store. While
this growth has supported data systems up till now, the size and complexity of
data being produced by surveys and even pointed telescope archives is reaching
scales where the requirements placed on file access by the \gls{POSIX} standard are
significantly hindering our ability to work with data.  Different parallel file systems
have different strengths and weaknesses.

At large scale, data service providers such as Dropbox and \gls{AWS} do not store files
in \gls{POSIX} systems.  Rather they present the illusion of directory structure layered over
large scale object stores. This allows for faster file access, with only \gls{CRUD} style
functions taking place on each object.  Further, as the pseudo-filesystem layer is simply a
view of structure typically provided by a graph database, users can arrange or potentially
have real time query driven structure for the file organization, removing how many now
organize data through a sea of nested symlinks.
Some provide local \gls{POSIX} caches of that users view of the
pseudo-filesystem, allowing for \gls{POSIX} style applications to access the files with
\texttt{fopen}, \texttt{fscan}, and \texttt{fclose} standard commands. Additionally, these providers do
not show users how data are stored. One can simply request data in the format
needed (e.g., Excel, \gls{CSV}, or as I assume it is the \gls{JSON} format the web apps
use for sheets).   At scale, applications often forgo
such POSIX layers and simply use the \gls{CRUD} interfaces to the objects to load them into
memory, act on the objects, and then update or delete them, in the format they need
as input and with the format they naively produce. Further, the data providers need
not update their data archive when formats change, simply provide a new updated
data access format that can be fueled by legacy data formats.

It is time for Astronomy data researchers to follow this curve. As users have migrated
from using tools on their laptops to support collaboration while reducing
individuals need to manage their systems (Jupyter Hubs, Overleaf, Google Slides),
so should astronomical data processing and analysis. We propose the adoption of
a common Astronomical data access \gls{API} layer.

\section{Recommendations }
\begin{enumerate}
 \item Develop a community wide architecture supporting Science as a Service following
	an industry  standard layered architecture for astronomy processing and data access as
depicted in \figref{fig:ci}.

\begin{figure}
\centering
\includegraphics[width=0.49\textwidth]{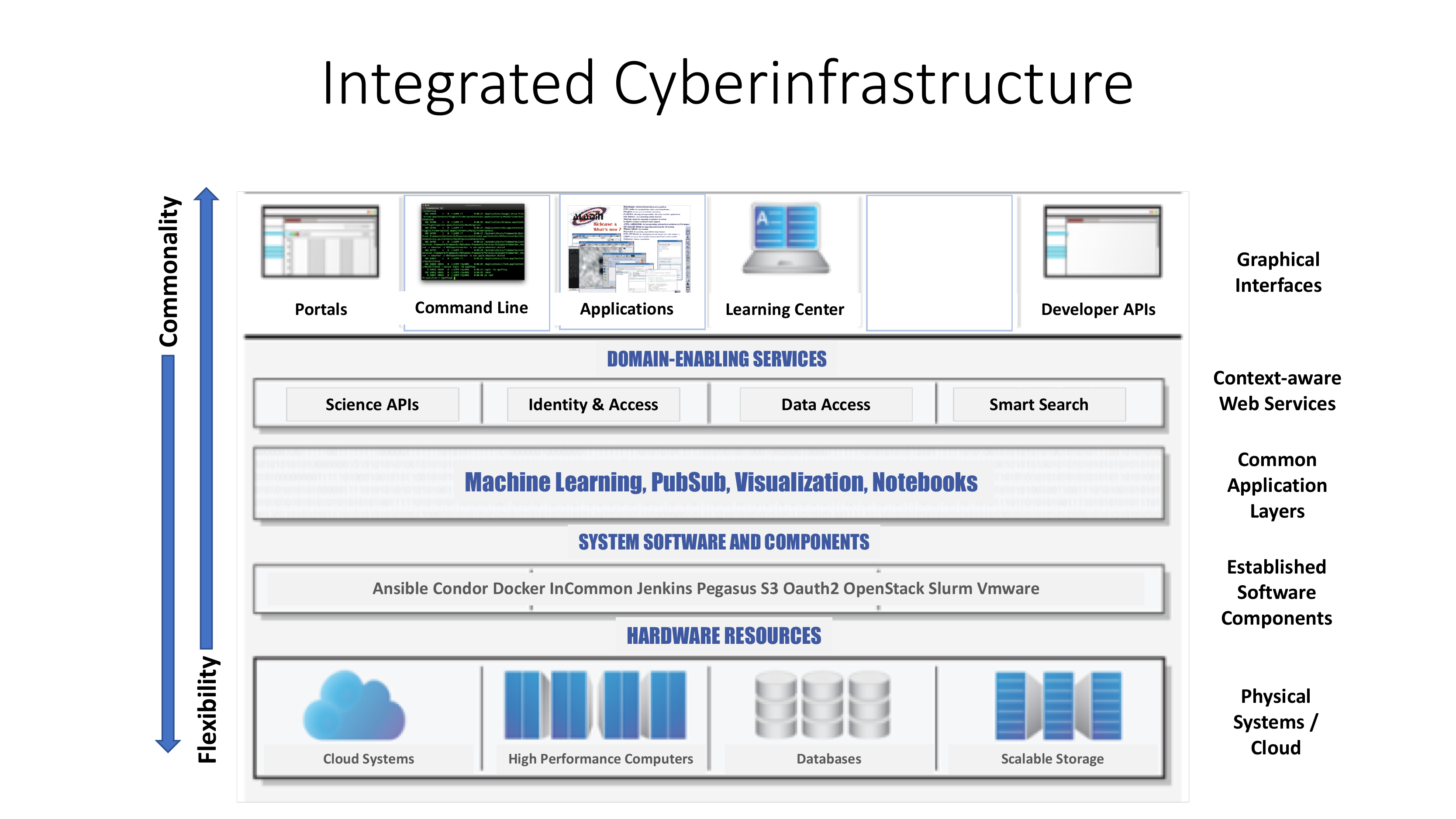}
\includegraphics[width=0.49\textwidth]{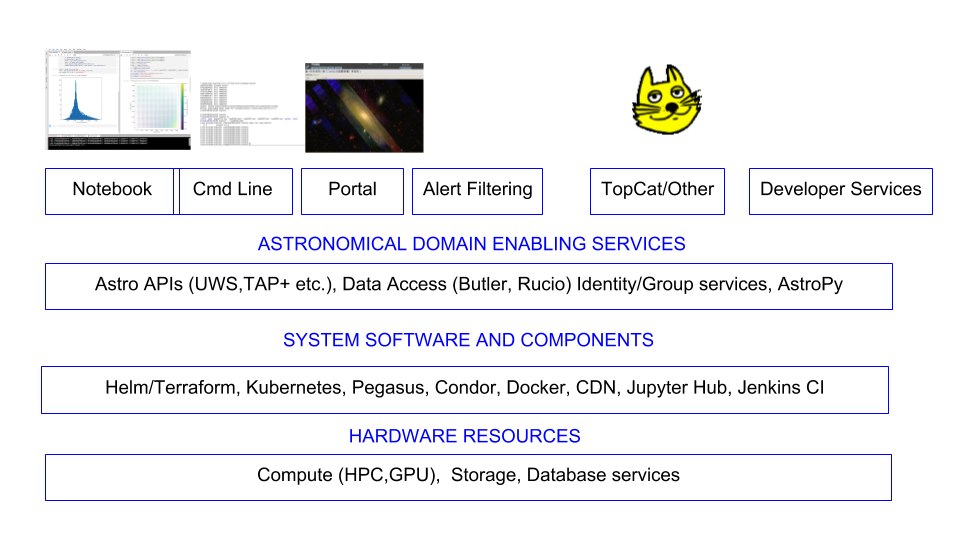}
\caption{Industry standard Cyber infrastructure model (left) and an astronomy instantiation of such a model (right)\label{fig:ci}}
\end{figure}

\item Compel all funded funded astronomical projects with \emph{software deliverables} to  leverage the architecture and  APIs for data access.

\item Develop \emph{data and \gls{metadata} transform services}
to support these common \gls{API} layers to aid interoperation.

\item Adopt and subscribe to a \emph{common federated identity service} (e.g., InCommon or
Globus Auth) that is supported by most university and research organizations.
\end{enumerate}

\section{Commodity services and software based community architecture} \label{sec:refarc}
The astronomy and astrophysics community have historically relied on the development and use of bespoke software and hardware infrastructure to solve challenges related to the managing and analyzing datasets at a scale that was difficult to find in industry or other scientific domains.
These requirements are no longer unique and we have access to a wealth of open source software, commodity hardware, and managed cloud services (offered by commercial providers and federally-funded institutions) that are well positioned to meet the needs of astronomers and astrophysicists \citep{2019AAS...23345706M, 2019AAS...23324505B}.
By providing documentation and reference implementations of the “astronomy \gls{stack}” using these technologies and making it easier for researchers and missions to access cloud computing services, we can reduce operations costs, accelerate time to science, and increase the scientific return on Federally-funded research in astronomy and astrophysics.

\figref{fig:ci} shows the
layers of the \gls{CI} from the interfaces for service access exposed
at multiple levels, the common domain wide enabled services, and a collection of system level components that support the
higher levels of the \gls{CI}.
The lower down the diagram are commodity layers based on well established and supported
components. As one moves up from these layers, more abstraction can be done to
expose these pieces in domain, or highly specific subdomain, level interfaces. By making these
abstractions, more universal service can be developed that can be applied more globally
across the entirety of the \gls{CI} as a whole.

An example of this would be
authentication, where each university or agency may provide their own authentication method
but unifying services like CILogon can bring those together to give global spaced
identity for a wide range of users based on disparate authentication systems.
By providing this structure along with a reference architecture of these System Services based on
well supported software components, providers are easily able to both deploy and support these common services which enable
cross mission and center interoperability. This structure also reflects how this architecture allows for greater reusability as one gets closer to the actual implementation of these
services while supporting greater flexibility and general usability as one works further from the core components.
Alternatives such as using github authentication may be more flexible but lack the
 rigor of InCommon which assures the individual is a member of an academic body --- we must also work with these industry standards though.

This service architecture should be based on using standard reusable software from many of the established standards developed outside of astronomy (e.g., common authentication mechanisms such as CILogon, standard data and \gls{metadata} management systems).  Standard \gls{API} interfaces should also be used to expose these components to higher level \gls{API}s. Data
formatting and \gls{metadata} structure can be exposed at the service level, allowing for
more data and \gls{metadata} reuse.
 An example of a storage agnostic data/metadata layer is the \gls{LSST} data butler \citep{2018arXiv181208085J}. Creation  has been driven by the need to abstract data access away from algorithms.
The \gls{algorithm} code deals with Python objects and never directly with data formats or even storage. The data butler is a lot more than a data access layer, it includes a full registry of all data objects and how to locate them. This means it may sit atop a filesystem or an object store --- a prototype S3 plugin is now available and we are testing it in a \gls{pipeline}.  The data butler is astronomy oriented --- it has a built in understanding of certain relationships such as between observations and calibrations or between observations and region of the sky. Since all metadate is held in a registry \gls{provenance} queries can be answered by the butler and it already can act as the registry to find objects in an object store.

Part of a Cyber Infrastructure model such as depicted in \figref{fig:ci} is an object store oriented \gls{API} --- this should be used for data sharing. Such APIs exist such as Amazon's S3.
The  \gls{API} layers must \emph{expose both data and \gls{metadata} in common
transferable and transformable formats} (e.g., \gls{CAOM}).

There must also be an authentication source for users at institutions without such means and for citizen science efforts.

\section{Why we should kill the filesystem}

Users should not care and repositories should not be tied to legacy formats  and storage representations because of legacy constraints  at other repositories.
The rest of the world has already moved on,  Google, Amazon, GitHub, Netflix etc. do not host large filesystems and and can scale because they are not limited by this antiquated formalism.

Filesystems with name spaces are very fragile at large scale. As we get larger data sets we have to trick the filesystem to not run out of Inodes, we make countless sub directories to cope with our thousands of files.
This is turn leads to countless hours spent fighting over how to organize files  the \emph{right way} in a filesystem.
Countless years have been spent fighting over data formats (\gls{FITS} vs \gls{HDF} \citep{2015ASPC..495...11M} vs \gls{CSV} vs Pandas).
If we move code then perhaps the filesystem is not organised in the same manner and the code may not work --- remote access to allow caching is not always an option.

We need to foster better remote collaboration.  The laptop is the bane of file sharing.
This has changed with cloud based pseudo-file systems but require storage in a single
cloud providers infrastructure. By creating a Filesystem as a Service (\gls{FSAAS}) federated
across data and cloud providers, we will win.

This would imply that \emph{POSIX based file access be deprecated
in software development} and only used when applications require thread safe
data access (something that is currently not possible with \gls{FITS} files).
We should however  develop a pseudo-directory structure system to
integrate local and remote files into a dynamic namespace for each user and potentially
each users use-case (e.g., the Box sync interface or the \gls{FUSE} based WholeTale file system
\citep{BRINCKMAN2019854})

This "Infrastructure as Code" \citep{morris2016infrastructure} approach lowers the bar to entry
and allows for easier adoption of more standardized services that will enable large-scale
astronomical research in ways that are well demonstrated in plant genomics (CyVerse and Galaxy), natural hazards (Designsafe), and surface water research (Hydroshare). (See also the decadal paper on Cloud infrastructure by Arfon Smith et. Al.)

\subsection{The catch }
Switching to an object store removes the filesystem bottle neck however it also removes the filesystem index. This implies that a registry of the objects must be contained. We frequently do this anyway usually sucking meta data into a databases to allow searching --- just in the case of an object store this would no longer be  optional.

\section{Conclusion}
We should agree on an \emph{astronomy \gls{stack}} of services with agreed interfaces such that we can concentrate on building domain specific layers on top of industry standard tools.

We should stop worrying about filesystems in astronomy --- we should agree on a decent \gls{API} for object storage and a registry to go with it.
The registry should build on existing agreements i.e., based on \gls{CAOM}-2 \citep{2007ASPC..376..347D}.

Adoption of  a limited set of  services will  aid ease of use and cut down on wasted effort by all data providers.

\bibliographystyle{yahapj}
\bibliography{local,lsst,lsst-dm,refs_ads,refs,books}

\begin{thebibliography}{}
\providecommand\natexlab[1]{#1}
\providecommand\JournalTitle[1]{#1}

\bibitem[{{Bektesevic} {et~al.}(2019){Bektesevic}, {Mehta}, {Juric}, {Slater},
  {Balazinska}, \& {Connolly}}]{2019AAS...23324505B}
{Bektesevic}, D., {Mehta}, P., {Juric}, M., {et~al.} 2019, in American
  Astronomical Society Meeting Abstracts, Vol. 233, American Astronomical
  Society Meeting Abstracts \#233, 245.05

\bibitem[{Brinckman {et~al.}(2019)Brinckman, Chard, Gaffney, Hategan, Jones,
  Kowalik, Kulasekaran, Ludäscher, Mecum, Nabrzyski, Stodden, Taylor, Turk, \&
  Turner}]{BRINCKMAN2019854}
Brinckman, A., Chard, K., Gaffney, N., {et~al.} 2019,
  \href{http://dx.doi.org/https://doi.org/10.1016/j.future.2017.12.029}{\JournalTitle{Future
  Generation Computer Systems}, 94, 854 }

\bibitem[{{Dowler} {et~al.}(2007){Dowler}, {Gaudet}, {Durand}, {Redman},
  {Hill}, \& {Goliath}}]{2007ASPC..376..347D}
{Dowler}, P.~D., {Gaudet}, S., {Durand}, D., {et~al.} 2007, in Astronomical
  Society of the Pacific Conference Series, Vol. 376, Astronomical Data
  Analysis Software and Systems XVI, ed. R.~A. {Shaw}, F.~{Hill}, \& D.~J.
  {Bell}, 347

\bibitem[{{Jenness} {et~al.}(2018){Jenness}, {Bosch}, {Schellart}, {Lim},
  {Salnikov}, \& {Gower}}]{2018arXiv181208085J}
{Jenness}, T., {Bosch}, J.~F., {Schellart}, P., {et~al.} 2018,
  \JournalTitle{arXiv e-prints}, arXiv:1812.08085

\bibitem[{{Mink} {et~al.}(2015){Mink}, {Mann}, {Hanisch}, {Rots}, {Seaman},
  {Jenness}, {Thomas}, \& {O'Mullane}}]{2015ASPC..495...11M}
{Mink}, J., {Mann}, R.~G., {Hanisch}, R., {et~al.} 2015, in Astronomical
  Society of the Pacific Conference Series, Vol. 495, Astronomical Data
  Analysis Software an Systems XXIV (ADASS XXIV), ed. A.~R. {Taylor} \&
  E.~{Rosolowsky}, 11

\bibitem[{{Momcheva} {et~al.}(2019){Momcheva}, {Smith}, \&
  {Fox}}]{2019AAS...23345706M}
{Momcheva}, I., {Smith}, A.~M., \& {Fox}, M. 2019, in American Astronomical
  Society Meeting Abstracts, Vol. 233, American Astronomical Society Meeting
  Abstracts \#233, 457.06

\bibitem[{Morris(2016)}]{morris2016infrastructure}
Morris, K. 2016, Infrastructure as Code: Managing Servers in the Cloud, Safari
  Books Online (O'Reilly Media)

\end{thebibliography}

\printglossaries
\end{document}